\documentclass[11pt]{article}
\usepackage[T1]{fontenc}
\usepackage[utf8]{inputenc}
\usepackage{lmodern}
\usepackage[margin=1in]{geometry}
\usepackage{amsmath,amssymb,amsthm,mathtools}
\usepackage{booktabs,array,enumitem,microtype}
\usepackage{xcolor,tikz}
\usepackage{todonotes}
\usepackage{float}
\usepackage{listings}
\usetikzlibrary{arrows.meta,positioning,calc,decorations.pathreplacing,shapes.geometric,matrix,backgrounds,fit}
\usepackage[colorlinks=true,linkcolor=blue!45!black,
 citecolor=blue!45!black,urlcolor=blue!45!black]{hyperref}
\usepackage{bookmark}
\hypersetup{pdftitle={The Permanent in O*((2-epsilon)\string^n) Time},
 pdfsubject={Dynamic programming for the permanent}}
\newtheorem{theorem}{Theorem}[section]
\newtheorem{lemma}[theorem]{Lemma}
\newtheorem{proposition}[theorem]{Proposition}
\newtheorem{corollary}[theorem]{Corollary}
\newtheorem{fact}[theorem]{Fact}
\theoremstyle{definition}
\newtheorem{definition}[theorem]{Definition}

\theoremstyle{remark}

\DeclareMathOperator{\per}{per}
\DeclareMathOperator{\supp}{supp}
\DeclareMathOperator{\pref}{pref}
\newcommand{\NN}{\mathbb N}
\newcommand{\ZZ}{\mathbb Z}
\newcommand{\cP}{\mathcal P}
\newcommand{\cR}{\mathcal R}

\newcommand{\floor}[1]{\left\lfloor #1\right\rfloor}
\newcommand{\ind}[1]{[#1]}
\newcommand{\entropy}{H}
\makeatletter
\newcommand{\dplabel}[1]{%
  \mathrlap{\text{\hspace*{-\eqnshift@}\textsc{#1}:}}}
\makeatother
\newcommand{\abparrow}[1]{\xrightarrow{\mathmakebox[2em]{#1}}}
\DeclarePairedDelimiterXPP{\Ostar}[1]{O^*\hspace{-0.02cm}}{(}{)}{}{#1}
\definecolor{smallstar}{RGB}{40,91,139}
\definecolor{deferred}{RGB}{174,108,29}
\definecolor{future}{RGB}{38,128,110}
\lstdefinelanguage{Lean}{
  morekeywords={theorem,example,def,by,exact,forall,fun,open,import,using,simpa,inductive,structure,where,noncomputable,if,then,else},
  morekeywords=[2]{ABP,Label,CommRing,Fin,Type,Equiv,Perm},
  sensitive=true,
  morecomment=[l]{--},
  morecomment=[s]{/-}{-/}
}
\lstdefinestyle{lean}{
  language=Lean,
  basicstyle=\ttfamily\footnotesize,
  keywordstyle=\color{smallstar}\bfseries,
  keywordstyle=[2]\color{future},
  commentstyle=\color{black!55},
  columns=fullflexible,keepspaces=true,showstringspaces=false,
  breaklines=true,xleftmargin=1em,xrightmargin=1em,
  aboveskip=.7\baselineskip,belowskip=.7\baselineskip,
  literate={ℝ}{{$\mathbb R$}}1 {ℕ}{{$\mathbb N$}}1 {ℤ}{{$\mathbb Z$}}1
    {∃}{{$\exists$}}1 {∀}{{$\forall$}}1 {∧}{{$\land$}}1
    {≤}{{$\le$}}1 {→}{{$\to$}}1
    {∑}{{$\sum$}}1 {∏}{{$\prod$}}1
}

\tikzset{
 diagram/.style={x=1cm,y=1cm,font=\footnotesize,>=Stealth},
 diagram line/.style={line width=.6pt,line cap=round,line join=round},
 diagram edge/.style={diagram line,draw=black!55},
 diagram vertex/.style={circle,draw=black!65,fill=white,line width=.6pt,
                        inner sep=0pt,minimum size=4.5pt,outer sep=1pt},
 diagram group/.style={fill=black!3,draw=black!20,line width=.4pt,
                       rounded corners=3pt},
 diagram heading/.style={font=\footnotesize,text=black!65},
 diagram label/.style={font=\footnotesize,fill=white,inner sep=1.5pt}
}

\tikzset{
 abp sequence/.style={diagram,
  execute at begin picture={\path (-1.4,0) -- (14.7,0);}},
 abp sequence label y/.initial=-3.02,
 abp sequence label anchor/.initial=north,
 abp sequence labels above/.style={abp sequence label y=.25,
  abp sequence label anchor=south},
 abp sequence block/.style={draw=black!60,fill=black!3,
  line width=.5pt,rounded corners=2pt},
 abp sequence transition/.style={diagram edge,draw=black,->},
 abp sequence weight/.style={midway,above=3pt,font=\scriptsize,
  text=black,inner sep=0pt},
 abp sequence rule/.style={midway,below=3pt,font=\scriptsize,
  text=black,inner sep=0pt},
 abp sequence edge/.style={diagram line,line width=.85pt},
 abp sequence closed edge/.style={abp sequence edge,draw=black},
 abp sequence open edge/.style={abp sequence edge,draw=smallstar},
 abp sequence deferred edge/.style={abp sequence edge,draw=deferred},
 abp sequence unused vertex/.style={fill=black!30,draw=black!30},
 abp sequence closed vertex/.style={fill=black,draw=black},
 abp sequence open vertex/.style={fill=smallstar,draw=smallstar},
 abp sequence deferred vertex/.style={fill=deferred,draw=deferred}
}

\newcommand{\abpSequenceEdgeSet}[2]{%
 \if\relax\detokenize{#2}\relax\else
  \foreach \leaf/\image in {#2} {
   \draw[abp sequence #1 edge] (u\leaf) -- (v\image);
   \path[abp sequence #1 vertex]
    (u\leaf) circle[radius=2pt] (v\image) circle[radius=2pt];
  }
 \fi
}

\newcommand{\abpSequenceSnapshot}[6][]{%
 \begin{scope}[xshift=#2cm]
  \if\relax\detokenize{#1}\relax\else
   \foreach \first/\last in {#1} {
    \path[abp sequence block]
     (-.49,{-.34*(\first-1)+.14}) rectangle
     (-.15,{-.34*(\last-1)-.14});
   }
  \fi
  \foreach \i in {1,...,9} {
   \coordinate (u\i) at (-.32,{-.34*(\i-1)});
   \coordinate (v\i) at (.32,{-.34*(\i-1)});
   \path[abp sequence unused vertex,line width=.5pt]
    (u\i) circle[radius=2pt] (v\i) circle[radius=2pt];
  }
  \abpSequenceEdgeSet{closed}{#3}
  \abpSequenceEdgeSet{deferred}{#5}
  \abpSequenceEdgeSet{open}{#4}
  \node[anchor=\pgfkeysvalueof{/tikz/abp sequence label anchor},
   font=\footnotesize,inner sep=0pt]
   at (0,\pgfkeysvalueof{/tikz/abp sequence label y})
   {$\begin{gathered}[t]#6\end{gathered}$};
 \end{scope}
}

\newcommand{\abpSequenceTransition}[4][]{%
 \draw[abp sequence transition,#1] ({#2+.6},-1.36) --
  node[abp sequence weight] {$#3$}
  node[abp sequence rule] {\textsc{#4}}
  ({#2+2.05},-1.36);
}

\setlist{topsep=4pt,itemsep=3pt}
\title{Get Permanent Results Faster by Counting Stars}
\author{Radu Curticapean\thanks{University of Regensburg and IT University of Copenhagen.
Funded by the European Union (ERC, CountHom, 101077083).
Views and opinions expressed are however those of the author(s) only and do not
necessarily reflect those of the European Union or the European Research
Council Executive Agency.}}
\date{}

\begin{document}
\maketitle
\begin{abstract}
We show that $n \times n$ permanents can be computed with $\Ostar{(2-\varepsilon)^n}$ arithmetic operations over every commutative ring, for $\varepsilon = 10^{-40}$. This solves a problem stated in Knuth's \emph{The Art of Computer Programming}.
Our algorithm implements the classical Binet--Minc identity with dynamic programming and careful bookkeeping.
It also gives algebraic branching programs of size $\Ostar{(2-\varepsilon)^n}$. 
\end{abstract}

\subsection*{AI Statement}
This work was developed in an exchange with GPT-6~Astra.
The author laid out the old story that embedding counts are linear combinations of homomorphism counts (see, e.g.,~\cite{lovasz,cdm,roth,roth-schmitt,drsw,rsw,cayley,fr,dmw,seppelt,neuen,agrs,cfi,curticapean-survey,roth-survey,afs}) and suggested finding a cutoff between different evaluation methods in view of the superexponential number of terms incurred by this technique.
The LLM developed the vast majority of the technical work and the author subsequently invested significant effort in distilling the core idea and improving the presentation.

\section{Introduction}\label{sec:overview}

We give a dynamic program that computes the permanent of an $n\times n$ matrix $A$
\[
                 \per(A)=\sum_{\sigma\in S_n}\prod_{i=1}^n A_{i,\sigma(i)}
\]
significantly faster than previous algorithms.
Below, let $\Ostar{\cdot}$ suppress polynomial factors.

\begin{theorem}\label{thm:main}
Let $\varepsilon = 10^{-40}$.
The permanent of an $n\times n$ matrix $A$ over any commutative ring with identity can
be computed in $\Ostar{(2-\varepsilon)^n}$ ring operations.
Moreover, there is an $\Ostar{(2-\varepsilon)^n}$ time algorithm that uniformly constructs an algebraic branching program for the $n\times n$ permanent in indeterminates $x_{ij}$. Its transition labels are among $x_{ij}$ and $0,1,-1$.
\end{theorem}

For comparison, known algorithms based on Ryser's and Glynn's formulas use $O(2^n n)$ operations~\cite{ryser,glynn}.
Knuth asked whether $\per(A)$ for a general \emph{real} $n\times n$ matrix $A$ can be computed using fewer than $2^n$ arithmetic operations~\cite[Exercise~4.6.4--11, p.~497]{knuth}.
Writing the running time as $O(2^n/s(n))$, Bj\"orklund~\cite{tabulation}
and Li~\cite{li} obtain savings $s(n)=\exp(\Omega(\sqrt{n/\log n}))$ and
$\exp(\Omega(\sqrt n))$, respectively, for \emph{integer matrices} with polynomially many bits.
For finite rings with $q\leq n$ elements, Bj\"orklund and Williams~\cite{bw} achieve savings $\exp(\Omega(n/q))$.
Knuth's question for general real matrices however remained unanswered,
and savings were only known \emph{conditionally}~\cite{bkkn} under Strassen's asymptotic rank conjecture~\cite{strassen-arc,conner-arc}.

\subsection{The Algorithm}

Throughout, we view $A$ as the edge weights of a complete bipartite graph $G=(L\cup R, E)$ with left vertices $L = \{u_1,\ldots,u_n \}$, right vertices $R = \{v_1,\ldots,v_n \}$, and weight $A_{ij}$ on the edge $u_i v_j$.
Moreover, we write $\cP_n$ for the set-partitions of $[n]$, with $|\cP_n|=\exp((1+o(1))n\log n)$.

\subsubsection*{The Binet--Minc identity}
\begin{figure}[t]
\centering
\begin{tikzpicture}[diagram]
 \node[diagram heading] at (.4,.65) {matching $M$};
 \node[diagram heading] at (6.0,.65) {star forest $M/\pi$};
 \node[diagram heading] at (11.65,.65) {image in $G$};

 \foreach \i in {1,...,6} {
  \pgfmathsetmacro{\yy}{-.5*(\i-1)}
  \ifnum\i<4\def\starcolor{deferred}\else
   \ifnum\i<6\def\starcolor{smallstar}\else\def\starcolor{future}\fi
  \fi
  \draw[diagram line,\starcolor] (-.25,\yy) -- (.95,\yy);
  \fill[\starcolor] (-.25,\yy) circle (1.7pt);
  \fill[\starcolor] (.95,\yy) circle (2.25pt);
  \node[anchor=east] at (-.37,\yy) {$u_\i$};
  \node[anchor=west] at (1.07,\yy) {$s_\i$};

  \fill[\starcolor] (5.25,\yy) circle (1.7pt);
  \node[anchor=east] at (5.13,\yy) {$u_\i$};
  \fill[\starcolor] (10.45,\yy) circle (1.7pt);
  \node[anchor=east] at (10.33,\yy) {$u_\i$};
  \node[diagram vertex] (image\i) at (12.25,\yy) {};
  \node[anchor=west] at (12.37,\yy) {$v_\i$};
 }
 \foreach \yy in {0,-.5,-1} {
  \draw[diagram line,deferred] (5.25,\yy) -- (6.65,-.5);
  \draw[diagram line,deferred] (10.45,\yy) -- (image2);
 }
 \foreach \yy in {-1.5,-2} {
  \draw[diagram line,smallstar] (5.25,\yy) -- (6.65,-1.75);
  \draw[diagram line,smallstar] (10.45,\yy) -- (image4);
 }
 \draw[diagram line,future] (5.25,-2.5) -- (6.65,-2.5);
 \draw[diagram line,future] (10.45,-2.5) -- (image4);
 \foreach \yy/\col in {-.5/deferred,-1.75/smallstar,-2.5/future}
  \fill[\col] (6.65,\yy) circle (2.25pt);
 \foreach \i/\col in {2/deferred,4/black}
  \fill[\col] (image\i) circle (2.25pt);

 \draw[diagram edge,->] (1.75,-1.25) --
  node[diagram label,above=5pt,align=center] {identify\\centers by $\pi$} (4.45,-1.25);
 \draw[diagram edge,->] (7.25,-1.25) --
  node[diagram label,above=5pt,align=center] {map\\centers by $\varphi$} (9.55,-1.25);
\end{tikzpicture}
\caption{Identifying centers of $M$ according to the partition
$\pi=\{\{1,2,3\},\{4,5\},\{6\}\}$ produces a star forest $M/\pi$, and a left-pinned map $\varphi$ sends the star centers to right vertices.}
\label{fig:matching-homomorphism}
\end{figure}
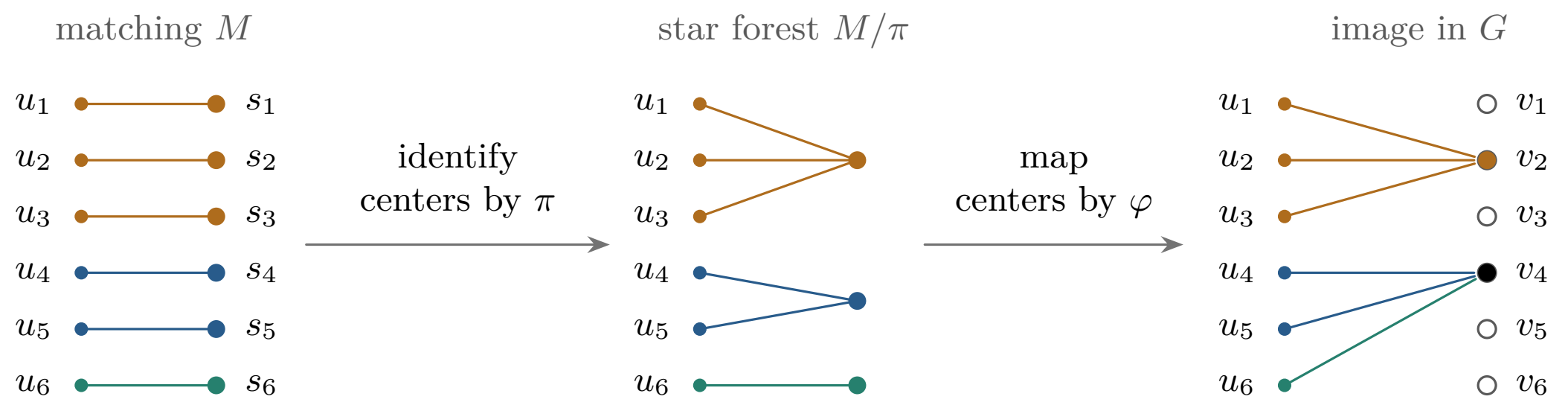

The starting point of our algorithm is the classical Binet--Minc formula, which expresses the permanent as a signed sum over $|\cP_n|$ terms that are individually easy to compute. 
As $|\cP_n|$ is superexponential, we cannot directly use this formula to obtain faster permanent algorithms.
In this paper, we develop mechanisms for evaluating the formula more efficiently.

To explain the Binet--Minc formula itself, start with the matching $M$ of $n$ disjoint edges $u_i s_i$.
A \emph{left-pinned map} $\varphi$ from $M$ to $G$ is a map $\varphi : V(M) \to V(G)$ with $\varphi(u_i) = u_i$ and $\varphi(s_i) \in R$ for all $i \in [n]$.
Its weight is $w(\varphi) = \prod_i A_{i,\varphi(i)}$, where we abuse notation and read $\varphi(s_i) = v_j$ as $\varphi(i) = j$.
Write $\Phi(M)$ for the set of left-pinned maps from $M$ to $G$.
As the \emph{injective} maps $\varphi\in\Phi(M)$ correspond to permutations of $[n]$, we have
\begin{equation}\label{eq:injective-maps}
 \per(A)=\sum_{\substack{\varphi\in\Phi(M)\\\text{injective}}}w(\varphi).
\end{equation}

While general maps $\varphi \in \Phi(M)$ need not be injective, we can isolate the injective maps by Möbius inversion.
For a partition $\pi \in \cP_n$, define the quotient $M/\pi$ of $M$ by identifying, for every block $B\in\pi$, the vertices $\{ s_i \mid i \in B \}$ to a vertex $s_B$. 
Then $M/\pi$ is a disjoint union of stars, a \emph{star forest}.
We define left-pinned maps from $M/\pi$ exactly as above and denote their set by $\Phi(M/\pi)$.
The Binet--Minc identity (cf.~Theorem~\ref{thm:Binet--Minc}) now recovers the right-hand side of~\eqref{eq:injective-maps}
by signed cancellation of map counts, using coefficients $\mu(\pi) \in \mathbb Z$:
\begin{equation}\label{eq:overview-expansion}
 \per(A)=\sum_{\pi\in\cP_n}\mu(\pi)
             \sum_{\varphi\in\Phi(M/\pi)}w(\varphi).
\end{equation}

For a fixed partition $\pi\in\cP_n$, the inner sum in \eqref{eq:overview-expansion} factors over the stars, as each star can independently choose a placement $\varphi(s_B) = v_j$ for its center $s_B$. We obtain:
\begin{equation}\label{eq:hom-count}
 \sum_{\varphi\in\Phi(M/\pi)}w(\varphi)
 =\prod_{B\in\pi}\left(\sum_{j=1}^n\prod_{i\in B}A_{ij}\right).
\end{equation}
The right-hand side can be evaluated with $O(n^2)$ ring operations.
Together with~\eqref{eq:overview-expansion}, this allows us to evaluate $\per(A)$ with $\Ostar{|\cP_n|}$ ring operations, which is far from optimal.

\subsubsection*{Dynamic programming baseline}
In preparation for the full construction, we first evaluate~\eqref{eq:overview-expansion} by a dynamic programming approach that constructs the partitions $\pi \in \cP_n$ and maps $\varphi\in\Phi(M/\pi)$ together, one star at a time. 
This already yields $\Ostar{2^n}$ time. 
At any given point in the construction, the program only remembers the set of left vertices used so far and a little information about one currently active star that is being constructed.
The full dynamic program, discussed later, will use more states.

More formally, the basic DP tracks contributions of partial maps that already cover a set $S\subseteq[n]$ of left vertices and have a currently active star of size $k \in \mathbb N$ whose center maps to a given right vertex $v_j \in R$. We use $(k,j)=(0,0)$ when no star is active. The table is organized as follows:
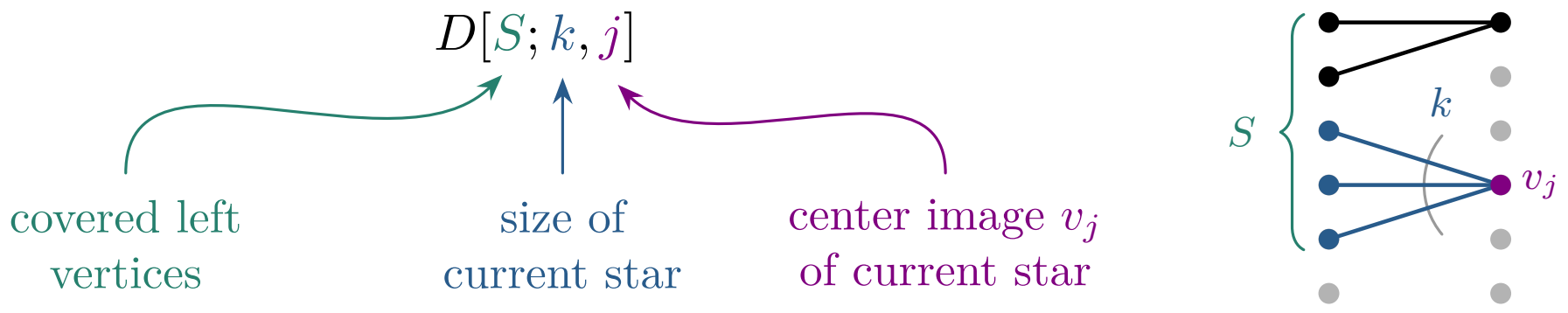
\begin{figure}[H]
\centering
\begin{tikzpicture}[diagram]
 \colorlet{coveredset}{future}
 \colorlet{centerimage}{violet}
 \matrix (indices) [matrix of math nodes,
  nodes={font=\normalsize,anchor=base,inner sep=0pt,outer sep=0pt},
  inner sep=0pt,column sep=0pt] {
  D[ & \textcolor{coveredset}{S} & {;\,} & \textcolor{smallstar}{k} & {,\,} & \textcolor{centerimage}{j} & ] \\
 };

 \node[font=\small,anchor=north,align=center,text=coveredset] (covered) at (-3.1,-1.1)
  {covered left\\vertices};
 \node[font=\small,anchor=north,align=center,text=smallstar] (size) at (indices-1-4.center |- covered.north)
  {size of\\current star};
 \node[font=\small,anchor=north,align=center,text=centerimage] (image) at (3.1,-1.1)
  {center image $v_j$\\of current star};

 \draw[diagram edge,draw=coveredset,->,shorten <=2pt,shorten >=2pt]
  (covered.north) to[out=90,in=225] ([yshift=-4pt]indices-1-2.south);
 \draw[diagram edge,draw=smallstar,->,shorten <=2pt,shorten >=2pt]
  (size.north) -- ([yshift=-4pt]indices-1-4.south);
 \draw[diagram edge,draw=centerimage,->,shorten <=2pt,shorten >=2pt]
  (image.north) to[out=90,in=-45] ([yshift=-4pt]indices-1-6.south);
 \begin{scope}[xshift=6cm,yshift=.11cm]
  \foreach \i in {1,...,6} {
   \coordinate (partial-u\i) at (0,{-.41*(\i-1)});
   \coordinate (partial-v\i) at (1.3,{-.41*(\i-1)});
   \path[abp sequence unused vertex,line width=.5pt]
    (partial-u\i) circle[radius=2pt]
    (partial-v\i) circle[radius=2pt];
  }

  \draw[diagram line,draw=black!40]
   ($(partial-v4)+(140:.58)$) arc[start angle=140,end angle=220,radius=.58];
  \node[text=smallstar,inner sep=0pt,yshift=4pt] at (.85,-.74) {$k$};

  \foreach \i in {1,2} {
   \draw[abp sequence closed edge] (partial-u\i) -- (partial-v1);
   \path[abp sequence closed vertex]
    (partial-u\i) circle[radius=2pt];
  }
  \path[abp sequence closed vertex] (partial-v1) circle[radius=2pt];
  \foreach \i in {3,4,5} {
   \draw[abp sequence open edge] (partial-u\i) -- (partial-v4);
   \path[abp sequence open vertex]
    (partial-u\i) circle[radius=2pt];
  }
  \path[fill=centerimage,draw=centerimage] (partial-v4) circle[radius=2pt];

  \draw[diagram line,draw=coveredset,decorate,
        decoration={brace,amplitude=5pt,mirror}]
   (-.19,.06) -- (-.19,-1.72)
   node[midway,left=10pt,text=coveredset,inner sep=0pt] {$S$};
  \node[anchor=west,text=centerimage,inner sep=0pt]
   at (1.45,-1.23) {$v_j$};
 \end{scope}
\end{tikzpicture}
\end{figure}

Starting from the initial entry $D[\varnothing;0,0]=1$, each step either
\begin{itemize}
 \item \textbf{begins} a new star with the smallest yet uncovered left vertex $u_i$
 and chooses its center image $v_j$ on the right, thereby proceeding to $D[S\cup\{i\};1,j]$, or
 \item \textbf{grows} the current star by adding an arbitrary leaf $i\notin S$ from the left vertices, or
 \item \textbf{closes} the current star, proceeding to $D[S;0,0]$.
\end{itemize}
The steps are weighted so that all ways of constructing $(\pi,\varphi)$ contribute a total of $\mu(\pi)w(\varphi)$ to the final value $D[[n];0,0]$, matching the contribution of $(\pi,\varphi)$ in~\eqref{eq:overview-expansion}.
Because the table has $\Ostar{2^n}$ entries and each depends on polynomially many earlier entries, this approach allows us to evaluate~\eqref{eq:overview-expansion} with $\Ostar{2^n}$ operations.
Our actual speedup from Theorem~\ref{thm:main} follows the same idea, but it adds an extra layer of bookkeeping.

\subsubsection*{If there were only small stars}

Before explaining the full DP leading to Theorem~\ref{thm:main}, imagine first that we could limit the stars grown by the basic DP to have at most $r$ leaves.
In this case, covering $|S|=s$ left vertices requires at least $s/r$ stars.
Since our DP rules force each new star to begin at the \emph{smallest} uncovered vertex, the set $S$ must contain the initial segment $1, \ldots, \lceil s/r\rceil$.
This forced prefix leaves only $2^{n-s/r}$ choices for $S$, which in turn leaves only $\Ostar{c_r^n}$ table entries for some $c_r<2$.
This idea was already used by Koivisto for partitioning into sets of bounded size~\cite[Section~2]{koivisto}.

\subsubsection*{Large stars may form in stellar nurseries}

While it would be convenient to simply discard large stars, we must find a way to handle them.
Our solution is to \emph{defer} their formation until the very end, where they can be handled in one batch. 

More concretely, for a cutoff $r \in \mathbb N$ chosen later, we modify the basic DP by (i) growing only stars with at most $r$ leaves, as above, and (ii) allowing the DP to defer the next vertex $u \in L$ to be destined for a large star. Whenever this happens, the DP commits to a right vertex $v \in R$ adjacent to $u$, but \emph{not} to the actual star with center $v$ that will receive $u$; recall that $\varphi$ may map several star centers to $v$.
At the end, we compute the contributions of all large stars simultaneously by summing over all groupings of leaves with the same center image into large stars.

This last step requires some bookkeeping, which we keep in a \emph{stellar nursery} $\mathbf d=(d_1,\ldots,d_n)$, where $d_j$ simply counts deferred leaves with center $v_j$.
This data suffices to reconstruct the relevant number of groupings, and it gives rise to an enlarged DP table
\[
D[S,\mathbf d;k,j].
\]

A crude bound suggests there could be up to $\binom{2n}{n}$ vectors $\mathbf d$, erasing any savings.
However, on closer inspection, a delicate trade-off arises:
As in the basic DP, each deferred leaf is chosen as the smallest uncovered vertex and therefore extends the initial segment covered by $S$.
This reduces the number of relevant DP states, exactly as in the small-star case discussed before.
Thus, while tracking stellar nurseries incurs an exponential cost, it also reduces the number of DP states by an exponential factor. 
The final speedup hinges on the balance between these two factors: the total number of DP states must remain bounded by $(2-\varepsilon)^n$.

\subsubsection*{From stargazing to number crunching}

To make this trade-off favorable, we additionally allow a budget of at most $\alpha_r n$ stars larger than $r$, for sufficiently small $\alpha_r>0$.
The centers of these stars then have at most $\alpha_r n$ distinct images, so we only need to consider stellar nurseries with at most $\alpha_r n$ nonzero entries.
Counting them together with the compatible covered sets $S$ yields an exponential saving.

To cover all partitions $\pi$, we run the DP at several cutoffs $r$ and large-star budgets $\alpha_r$.
As $r$ increases, more stars can be grown directly and fewer need the nursery.
By an elementary counting argument, we can identify a fixed finite list of geometrically increasing cutoffs and corresponding fractions $\alpha_r$, independent of $n$, such that \emph{every} partition $\pi$ is handled by some cutoff: 
The main text shows an existence statement, and the appendix shows that $280$ cutoffs suffice for Theorem~\ref{thm:main}.

To conclude, we assign each partition $\pi \in \cP_n$ to a \emph{unique} cutoff from the finite parameter list and remember this choice when forming large stars from the stellar nursery.
Summing contributions over the different cutoffs makes each $(\pi,\varphi)$ contribute exactly $\mu(\pi)w(\varphi)$.
Overall, this evaluates~\eqref{eq:overview-expansion} with $\Ostar{(2-\varepsilon)^n}$ operations for a fixed $\varepsilon>0$ and proves Theorem~\ref{thm:main}.

\subsection{Related Ideas}

The Binet--Minc identity generalizes formulas found by Binet in 1812; its general form was given by Minc in 1979~\cite{minc}.
It is an application of M\"obius inversion on the lattice of set partitions, which was used for more general subgraph counting problems~\cite{lovasz,cdm}.
The concrete Binet--Minc identity was used algorithmically for rectangular permanents~\cite{kt}.
It also occurs in circuit homogenization~\cite{lst-original},
set-multilinearization~\cite{forbes}, and in works on the complexity of monomial
symmetric polynomials~\cite{cls,bcdj}.

Our algorithm is inspired by Koivisto's
smallest-element ordering for set partitions with bounded block
size~\cite{koivisto}, and by the chromatic-number
algorithm of Bj\"orklund et al.~\cite{bchkp}, which combines
specialized coloring routines with three-way partitioning
under the asymptotic rank conjecture.

\subsection{Neither False nor Obvious}

An exponential improvement over the long-standing $\Ostar{2^n}$ time algorithms for permanents, under a dubious paper title, should naturally raise doubts.
One answer to such concerns is provided in the ancillary proof of the following Lean statement.
See Appendix~\ref{app:lean} for the definitions used.
\label{lst:lean-main}\begin{lstlisting}[style=lean,basicstyle=\ttfamily\small,xleftmargin=0pt,xrightmargin=0pt]
theorem main_abp :
  ∃ (C : ℝ) (d : ℕ), 0 < C ∧ ∀ n : ℕ, ∃ G : ABP n,
    (∀ A : Fin n → Fin n → ℤ, G.output A = permanent A) ∧
    (G.vertices + G.transitions : ℝ) ≤ C * (n+1 : ℝ)^d * (2 : ℝ)^((0.999994 : ℝ)*n)
\end{lstlisting}

This certifies that the ABP existence part (over $\mathbb Z$) asserted by Theorem~\ref{thm:main} \emph{is} true. 
Let us also explain why it \emph{might} be true, i.e., why it evades known lower bounds in restricted models.
\begin{enumerate}
      \item It is known that \emph{monotone} arithmetic circuits for the permanent require $\Omega(2^n)$ size~\cite{jerrum-snir}. Our algorithm however crucially uses signed terms and their cancellations.
      \item ABPs for the permanent require size $\Omega(2^n)$ when every source-to-sink path uses the same fixed row order~\cite[Theorem~2.4]{ikenmeyer-landsberg}. While our rule for beginning stars would be compatible with the ascending row order, arbitrary uncovered left vertices may be added as leaves to the current star. This means that different walks through the ABP can use different row orders.
      \item The standard conversion from ABPs to determinants translates our Theorem~\ref{thm:main} into a determinantal representation of dimension $\Ostar{(2-\varepsilon)^n}$ for the permanent. Landsberg and Ressayre~\cite{landsberg-ressayre} prove a $2^n-1$ lower bound under equivariance, but this requires that row permutations of the input lift to compatible symmetries of the representation. Because we need to choose the \emph{smallest} uncovered left vertex when beginning a new star, our rules impose an order on the left vertices, so our representation lies outside this model.
\end{enumerate}

Note also that the recent subexponential \#3-SAT algorithms~\cite{anthropic-eth,openai-eth}
give $2^{o(v+e)}$ time for counting perfect matchings in graphs with
$v$ vertices and $e$ edges, when used with a standard reduction. 
For dense $n\times n$ permanent instances, this
encoding has $\Theta(n^2)$ variables, so it does not give the bound
in Theorem~\ref{thm:main}.
However, a more efficient reduction is not ruled out.

\section{Preliminaries}\label{sec:preliminaries}

Throughout, rings are commutative with identity, and additions and
multiplications have unit cost.
For a statement $\phi$, the Iverson bracket $\ind{\phi}$ is $1$ if
$\phi$ holds and $0$ otherwise. 

The permanent counts injective left-pinned maps from the matching $M$ to $G$. Binet--Minc recovers this count from left-pinned maps of the star
forests $M/\pi$.
Identify each left vertex $u_i$ with its ordered label $i$.
For a nonempty block $B$ and a partition $\pi$, let
\begin{equation}\label{eq:mobius-coefficient}
 \mu(B)=(-1)^{|B|-1}(|B|-1)! \quad \text{and}\quad 
 \mu(\pi)=\prod_{B\in\pi}\mu(B).
\end{equation}

\begin{theorem}[Binet--Minc {\cite[Theorem~2]{minc}}]\label{thm:Binet--Minc}
Over every commutative ring with identity,
\begin{equation}\label{eq:Binet--Minc}
 \per(A)=\sum_{\pi\in\cP_n}\mu(\pi)\sum_{\varphi\in\Phi(M/\pi)}w(\varphi).
\end{equation}
\end{theorem}

\begin{proof}
Fix a map $f:[n]\to[n]$ sending each left vertex $u_i$ to the
right vertex $v_{f(i)}$, and let $\sigma$ partition the left vertices
by their images under $f$. The map $f$ contributes the weight
$\prod_{i=1}^n A_{i,f(i)}$ for each partition $\pi$ on whose blocks
it is constant, that is, for each $\pi$ refining $\sigma$.
The total coefficient of this contribution is therefore
\[
 \sum_{\pi\leq\sigma}\mu(\pi)
 =\prod_{C\in\sigma}
       \left(\sum_{\rho\in\cP(C)}\prod_{B\in\rho}\mu(B)\right).
\]
The inner sum is the signed sum over permutations of $C$: their cycles
have supports $B$, and a support of size $b$ admits $(b-1)!$ cycles,
each of sign $(-1)^{b-1}$. This sum equals $\ind{|C|=1}$, since for
$|C|>1$ composing with a fixed transposition pairs opposite signs. Exactly the bijections $f$ survive, each with coefficient one.
\end{proof}

The individual map counts in \eqref{eq:Binet--Minc} are easy to evaluate by
\eqref{eq:hom-count}, but there are superexponentially many partitions $\pi$.
We therefore construct $\pi$ and $\varphi$ together and share work whenever partial constructions have the same possible continuations. The following model captures such computations.

\begin{definition}\label{def:abp}
An \emph{algebraic branching program} (ABP) in a set of indeterminates $X$ is a tuple
$\mathcal B=(H,\mathsf s,\mathsf t,\lambda)$, where $H=(V,E)$ is a
finite acyclic digraph, $\mathsf s,\mathsf t\in V$ are source and sink nodes with no incoming and no outgoing transitions, respectively, and $\lambda$ assigns a polynomial in $X$ to each $e\in E$.
We call the elements of $V$ \emph{nodes} and those of $E$ \emph{transitions}.
For each $\tau\in V$, define
\begin{equation}\label{eq:abp-path-sum}
D_{\mathcal B}[\tau]=\sum_{P : \text{path }\mathsf s\leadsto\tau\text{ in }H}\operatorname{wt}(P)
\qquad 
\text{with }
\operatorname{wt}(P)=\prod_{e\in P}\lambda(e).
\end{equation}
The ABP computes $D_{\mathcal B}[\mathsf t]$ and we usually omit the subscript $\mathcal B$ from $D_{\mathcal B}$. 
Its size is $|V|+|E|$ and its \emph{effective size} counts the nodes reachable from $\mathsf s$. 
\end{definition}

With $D[\mathsf s]=1$, all node values can be evaluated by dynamic programming in topological order.
Joining disjoint ABPs to a new common source and sink by unit transitions
gives an ABP computing the sum of their outputs.

\section{Basic Dynamic Programming}\label{sec:baseline}

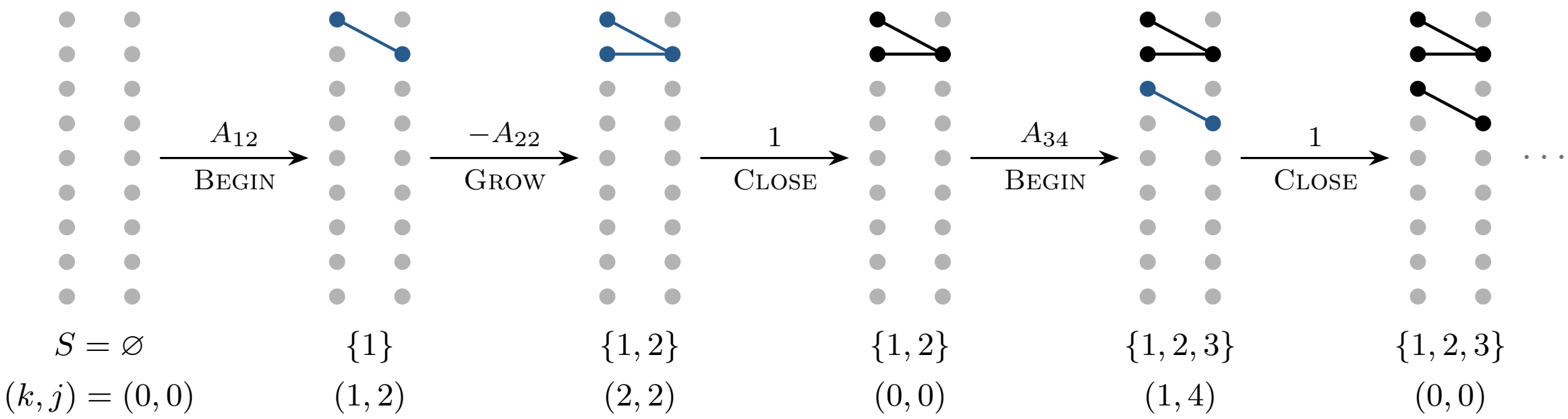
\begin{figure}[!htbp]
\centering
\begin{tikzpicture}[abp sequence]
 \abpSequenceSnapshot{0}{}{}{}
  {S=\varnothing\\(k,j)=(0,0)}
 \abpSequenceSnapshot{2.65}{}{1/2}{}
  {\{1\}\\(1,2)}
 \abpSequenceSnapshot{5.3}{}{1/2,2/2}{}
  {\{1,2\}\\(2,2)}
 \abpSequenceSnapshot{7.95}{1/2,2/2}{}{}
  {\{1,2\}\\(0,0)}
 \abpSequenceSnapshot{10.6}{1/2,2/2}{3/4}{}
  {\{1,2,3\}\\(1,4)}
 \abpSequenceSnapshot{13.25}{1/2,2/2,3/4}{}{}
  {\{1,2,3\}\\(0,0)}
 \abpSequenceTransition{0}{A_{12}}{Begin}
 \abpSequenceTransition{2.65}{-A_{22}}{Grow}
 \abpSequenceTransition{5.3}{1}{Close}
 \abpSequenceTransition{7.95}{A_{34}}{Begin}
 \abpSequenceTransition{10.6}{1}{Close}
 \node[text=black!55,font=\normalsize] at (14.2,-1.36) {$\cdots$};
\end{tikzpicture}
\caption{A path in the basic ABP for $n=9$ starts at the empty state. Blue edges belong to the open
star and black edges to closed stars; unused vertices are gray.
The path continues in Figure~\ref{fig:stellar-nursery-sequence}.}
\label{fig:abp-sequence}
\end{figure}

Following the introduction, we first realize the Binet--Minc sum \eqref{eq:Binet--Minc} by an ABP
whose paths construct star forests $M/\pi$ together with their left-pinned maps $\varphi$.
The ABP nodes are indexed by $(S;k,j)$, where $S\subseteq[n]$ tracks the covered left vertices, including the leaves of the currently open star. 
If $(k,j)=(0,0)$, then no star is open.
Otherwise, $k\in\{1,\ldots,n\}$ is the current number of leaves in the open star, and $j\in[n]$ specifies its center image $v_j$. 
Accordingly, the source and sink are $\mathsf s=(\varnothing;0,0)$ and $\mathsf t=([n];0,0)$. 
The construction rules sketched in the introduction manifest themselves in the ABP as three types of labeled transitions, see also Figure~\ref{fig:abp-sequence}:
\begin{alignat}{3}
 &\dplabel{Begin}
 &(S;0,0)
   &\abparrow{A_{ij}}(S\cup\{i\};1,j),
 &\qquad i&=\min([n]\setminus S),\quad j\in[n];
 \label{eq:base-start}\\
 &\dplabel{Grow}
 &(S;k,j)
   &\abparrow{-A_{vj}}(S\cup\{v\};k+1,j),
 &\qquad v&\notin S;
 \label{eq:base-grow}\\
 &\dplabel{Close}
 &(S;k,j)&\abparrow{1}(S;0,0).
 &&\label{eq:base-close}
\end{alignat}
Note that \textsc{Begin} requires a closed state with $S\ne[n]$, while
\textsc{Grow} and \textsc{Close} require an open star.
The ABP has at most $\Ostar{2^n}$ nodes, each with $O(n)$ outgoing
transitions: a closed node has at most $n$ \textsc{Begin} transitions, and an open
node has at most $n$ \textsc{Grow} transitions and one \textsc{Close} transition.

To obtain a topological order, sort the nodes by increasing $|S|$, with open states before closed
states at each level. Then \textsc{Close} advances within a level, and every other transition increases $|S|$.

\begin{lemma}\label{lem:baseline}
The ABP above computes $\per(A).$
\end{lemma}
\begin{proof}
Every source-to-sink path determines a partition $\pi\in\cP_n$ and
a left-pinned map $\varphi\in\Phi(M/\pi)$:
each \textsc{Begin} transition begins a star $B$ and chooses
$\varphi(s_B)=v_j$, the subsequent \textsc{Grow} transitions add its
remaining leaves, and \textsc{Close} ends the star.
The stars appear in order of their smallest leaves, and each star's
smallest leaf comes first. Conversely, a pair $(\pi,\varphi)$ together
with an ordering of each star's leaves with its smallest leaf first
determines a unique source-to-sink path.

Fix $(\pi,\varphi)$. For a star $B$ of size $b$ with
$\varphi(s_B)=v_j$, each of its
$(b-1)!$ leaf orders contributes the factor
$(-1)^{b-1}\prod_{i\in B}A_{ij}$ to the path weight. Summing these
factors over the leaf orders gives $\mu(B)\prod_{i\in B}A_{ij}$.
The leaf orders are chosen independently for each star, so the
paths realizing $(\pi,\varphi)$ have total weight
$\mu(\pi)w(\varphi)$. Summing over
$\varphi\in\Phi(M/\pi)$ gives $\mu(\pi)\sum_{\varphi\in\Phi(M/\pi)}w(\varphi)$,
and then summing over $\pi\in\cP_n$ gives $\per(A)$ by
\eqref{eq:Binet--Minc}.
\end{proof}

\section{Constructing the ABP}\label{sec:small}

For our main construction, we build on the ABP from Section~\ref{sec:baseline}.
For a cutoff $r\geq1$, the \emph{main part} described in Section~\ref{sec:deferred} grows stars with at most $r$ leaves and records deferred leaves in a \emph{stellar nursery}.
In Section~\ref{subsec:star-formation}, we describe \emph{star-formation gadgets}  that form the large stars from the stellar nursery post-hoc to obtain the \emph{combined ABP}.

The final construction relies on a specific parameter choice that is supplied in Section~\ref{sec:parameters}, which also bounds the effective size of the combined ABP. This section keeps these parameters unspecified for now and focuses on the ABP construction and correctness.

\subsection{Only Small Stars}
\label{sec:small-stars}

As a warm-up, consider a restricted Binet--Minc sum over pairs $(\pi,\varphi)$ such that all blocks in $\pi$ are of size at most $r$, i.e., all stars in $M/ \pi$ have at most $r$ leaves:
\begin{equation}\label{eq:bounded-sum}
 \sum_{\substack{\pi\in\cP_n\\\forall B\in\pi:\,|B|\leq r}}
       \mu(\pi)\sum_{\varphi\in\Phi(M/\pi)}w(\varphi).
\end{equation}

Koivisto's strategy~\cite{koivisto} for partitioning into sets of bounded size allows us to evaluate~\eqref{eq:bounded-sum}
with exponential savings: Ordering stars by their \emph{smallest leaves} forces a large covered prefix of left vertices, which significantly restricts the state space.

More concretely, we construct a modified ABP for~\eqref{eq:bounded-sum}
by keeping \textsc{Grow} transitions from the basic ABP only when $k<r$.
The source-to-sink paths in this modified ABP construct exactly the pairs
$(\pi,\varphi)$ with $|B|\leq r$ for every $B\in\pi$, and as in the proof of Lemma~\ref{lem:baseline}, the paths realizing each such pair
still have total weight $\mu(\pi)w(\varphi)$.
To bound the effective size of this ABP, consider a path from the source to a node $(S;k,j)$.
For $S\subseteq[N]$, define the prefix length
\begin{equation}\label{eq:pref}
 \pref(S)=\max\bigl\{p\in\{0,\ldots,N\}:[p]\subseteq S\bigr\}.
\end{equation}

Now say that the path to $(S;k,j)$ uses exactly $b \in \mathbb N$ transitions of type \textsc{Begin}. 
Each such \textsc{Begin} transition covers the smallest uncovered left vertex, so $\pref(S)\geq b$.
Moreover, each star has at most $r$ leaves, so $|S|\leq rb$.
Combining this, we find that only nodes with $|S|\leq r\pref(S)$ are reachable from the source in the modified ABP.
Thus the modified ABP has effective size $\Ostar{c_r^n}$ for some
$c_r<2$, later quantified in Lemma~\ref{lem:subset}.

\subsection{Stellar Nurseries}\label{sec:deferred}
\begin{figure}[t]
\centering
\begin{tikzpicture}[abp sequence,
 star-formation label/.style={font=\scriptsize,text=black,inner sep=0pt},
 curved weight/.style={star-formation label,midway,sloped,above=4pt},
 curved rule/.style={star-formation label,midway,sloped,below=4pt},
 sink weight/.style={star-formation label,pos=.55,right=4pt}]
 \node[text=black!55,font=\normalsize] at (-.95,-1.36) {$\cdots$};
 \abpSequenceSnapshot{0}{1/2,2/2,3/4}{}{}
  {S=\{1,2,3\}\\\mathbf d=\mathbf0}
 \draw[abp sequence transition,dashed] (.6,-1.36) --
  node[abp sequence weight] {$\prod_{i=4}^9 A_{i7}$}
  node[abp sequence rule] {\textsc{$6\times$ Defer}}
  (3.2,-1.36);
 \begin{scope}[local bounding box=star-formation-region]
  \node[diagram heading,anchor=north west,inner sep=0pt]
   at (3.45,2.5) {star-formation gadget};
 \abpSequenceSnapshot{3.8}{1/2,2/2,3/4}{}{4/7,5/7,6/7,7/7,8/7,9/7}
  {\{1,\ldots,9\}\\\mathbf d=6\mathbf e_7}
 \begin{scope}[abp sequence labels above]
  \abpSequenceSnapshot{6.45}{1/2,2/2,3/4}{4/7,5/7,6/7}{7/7,8/7,9/7}
   {(a;k,\nu_2)=\\(3;3,1)}

  \begin{scope}[yshift=2.25cm]
   \abpSequenceSnapshot{9.6}{1/2,2/2,3/4}{4/7,5/7,6/7,7/7}{8/7,9/7}
    {(2;4,1)}
   \abpSequenceSnapshot[4/9]{12.8}{1/2,2/2,3/4,4/7,5/7,6/7,7/7,8/7,9/7}{}{}
    {(0;0,1)}
   \draw[abp sequence transition,dashed] (10.2,-1.36) --
    node[abp sequence weight] {$1$}
    node[abp sequence rule] {\textsc{\shortstack{$2\times$ Grow\\Close}}}
    (12.2,-1.36);
  \end{scope}
  \begin{scope}[yshift=-2.25cm]
   \abpSequenceSnapshot[4/6]{9.6}{1/2,2/2,3/4,4/7,5/7,6/7}{}{7/7,8/7,9/7}
    {(3;0,1)}
   \abpSequenceSnapshot[4/6,7/9]{12.8}{1/2,2/2,3/4,4/7,5/7,6/7,7/7,8/7,9/7}{}{}
    {(0;0,2)}
   \draw[abp sequence transition,dashed] (10.2,-1.36) --
    node[abp sequence weight] {$1$}
    node[abp sequence rule] {\textsc{\shortstack{Begin\\$2\times$ Grow\\Close}}}
    (12.2,-1.36);
  \end{scope}
 \end{scope}
 \node[font=\footnotesize,inner sep=3pt] (star-formation-sink)
  at (14.45,-1.36) {$\mathsf t_r$};

 \abpSequenceTransition[dashed]{3.8}{1}{\shortstack{Begin\\$2\times$ Grow}}
 \draw[abp sequence transition] (7.05,-1.36)
  .. controls (7.75,-1.36) and (8.3,.89) ..
  node[curved weight] {$-1$}
  node[curved rule] {\textsc{Grow}}
  (9.0,.89);
 \draw[abp sequence transition] (7.05,-1.36)
  .. controls (7.75,-1.36) and (8.3,-3.61) ..
  node[curved weight] {$1$}
  node[curved rule] {\textsc{Close}}
  (9.0,-3.61);
 \draw[abp sequence transition,dashed] (13.4,.89)
  .. controls (14.25,.89) and (14.45,-.7) ..
  node[sink weight] {$1$} (star-formation-sink.north);
 \draw[abp sequence transition,dashed] (13.4,-3.61)
  .. controls (14.25,-3.61) and (14.45,-2.02) ..
  node[sink weight] {$1$} (star-formation-sink.south);
 \end{scope}
 \begin{pgfonlayer}{background}
  \node[diagram group,draw=none,fit=(star-formation-region),inner sep=4pt] {};
 \end{pgfonlayer}
\end{tikzpicture}
\caption{
\label{fig:stellar-nursery-sequence}  
Continuation of Figure~\ref{fig:abp-sequence} for cutoff $2$, with six orange leaves deferred to $v_7$.
The internal nodes of the star-formation gadget show $(a;k,\nu_2)$, blue edges belong to the open star,
and dashed arrows abbreviate paths, with unit advances through empty
clusters suppressed.
The fork shows representative paths to one six-leaf star or two triples;
summing all parallel \textsc{Grow} choices gives total weights
$-5!=-120$ and $\binom63/2\cdot(2!)^2=40$, respectively.}
\end{figure}
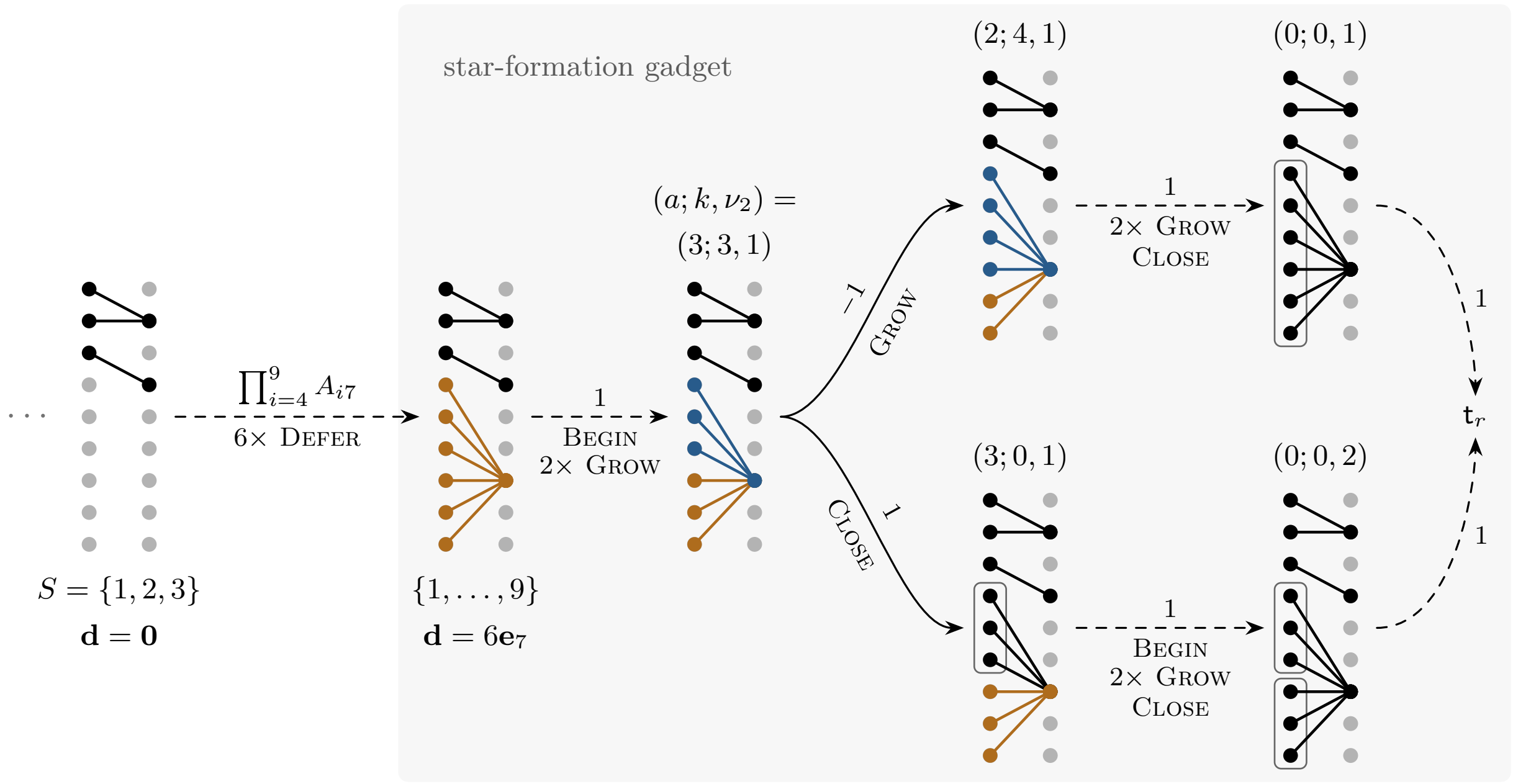

Next, we fix a cutoff $r$ as above and a rational fraction $0\leq\alpha_r\leq1$,
and allow for up to $\alpha_r n$ larger stars.
For a partition $\pi$, write $q_s(\pi)$ for the number of blocks $B \in \pi$ with $|B|> s$.
We first construct the main part for the restricted Binet--Minc sum
\begin{equation}\label{eq:large-star-sum}
 \sum_{\substack{\pi\in\cP_n\\q_r(\pi)\leq\alpha_r n}}
       \mu(\pi)\sum_{\varphi\in\Phi(M/\pi)}w(\varphi).
\end{equation}
In Section~\ref{subsec:star-formation}, we will further restrict this sum by assigning each partition to one cutoff.

The main part of the ABP keeps the transitions from Section~\ref{sec:small-stars},
with \textsc{Grow} still restricted to $k<r$.
For a leaf $i$ destined for a large star $B$, we choose its center image
$\varphi(s_B)=v_j$ but defer its block membership to be handled by a \emph{star-formation gadget} introduced in the next subsection.

To enable this strategy, we extend the indexing of the ABP nodes to $(S,\mathbf d;k,j)$, where $\mathbf d=(d_1,\ldots,d_n)\in\NN^n$ is the \emph{stellar nursery}.
Here $d_j$ counts deferred leaves with chosen center image $v_j$;
the star-formation gadget will use this information to count large stars.
For $S\ne[n]$, we maintain stellar nursery information through the following transitions, where $\mathbf e_j$ is the $j$th unit vector:
\begin{alignat}{2}
 &\dplabel{Defer}
 &(S,\mathbf d;0,0)
 &\abparrow{A_{ij}}(S\cup\{i\},\mathbf d+\mathbf e_j;0,0)
 \quad \text{with }i=\min([n]\setminus S) \text{ and } j\in[n],
 \label{eq:defer}
\end{alignat}

Note that, for normalization purposes, we may only $\textsc{Defer}$ if no small star is actively being constructed.
We can safely keep only nodes with $|\supp(\mathbf d)|\leq\alpha_r n$. 

The source is $(\varnothing,\mathbf0;0,0)$ and
Section~\ref{subsec:star-formation} will attach a star-formation gadget
at each node $([n],\mathbf d;0,0)$ of the main part,
leading to the common sink $\mathsf t_r$.
Figure~\ref{fig:stellar-nursery-sequence} summarizes the construction.

\subsection{Forming Stars in Stellar Nurseries}\label{subsec:star-formation}

Fix a finite set $\cR=\{r_1<\cdots<r_\ell\}$ of cutoffs and
rational fractions $0\leq\alpha_s\leq1$ for $s\in\cR$, all independent of $n$.
Their values will later be chosen in Section~\ref{sec:parameters}.

\begin{definition}\label{def:cutoff-assignment}
A cutoff $s\in\cR$ is \emph{admissible} for $\pi$ if $q_s(\pi)\leq\alpha_s n$.
We assume every partition of $[n]$, and hence (by singleton padding) every
partition of a subset of $[n]$, has an admissible cutoff.
We \emph{assign} each partition $\pi$ to its largest admissible cutoff $r(\pi)$.
\end{definition}

\begin{fact}\label{fact:large-block-assignment}
Let $\pi$ be a partition of a subset of $[n]$.
For $\rho=\{B\in\pi:|B|>r\}$, we have $r(\pi)=r$ if and only if $r(\rho)=r$.
This follows because $q_s(\pi)=q_s(\rho)$ for every $s\in\cR$ with $s\geq r$.
\end{fact}

We now clear the stellar nursery by forming large stars in it, essentially  re-using the DP from Section~\ref{sec:baseline}.
Consider a path to a closed terminal node
$\tau_{\mathbf d}=([n],\mathbf d;0,0)$ of the main part at cutoff $r$,
and let $U_j$ be its set of deferred leaves with chosen center image $v_j$, so $|U_j|=d_j$.
We call $U_j$ a cluster.
By Fact~\ref{fact:large-block-assignment}, it suffices to enforce
$r(\rho)=r$ for the partition $\rho$ formed by these large stars.

\begin{definition}[Star-formation gadget]\label{def:star-formation-gadget}
Given a stellar nursery $\mathbf d\in\NN^n$ with $\sum_jd_j\leq n$, choose disjoint clusters
$U_1,\ldots,U_n\subseteq[n]$ with $|U_j|=d_j$ and define
\begin{equation}\label{eq:star-formation-weight}
 C_r(\mathbf d)=\sum_{\rho:\,r(\rho)=r}\mu(\rho),
\end{equation}
where $\rho$ ranges over partitions of $U_1\cup\cdots\cup U_n$ into
blocks larger than $r$, each contained in a single cluster.
A \emph{star-formation gadget} for $\mathbf d$ is an ABP from
$\tau_{\mathbf d}$ to $\mathsf t_r$ computing $C_r(\mathbf d)$.
\end{definition}

\begin{lemma}\label{lem:star-formation-abp}
Every $\mathbf d$ admits a star-formation gadget of size
$O(n^{|\cR|+4})$ with labels $\pm1$.
\end{lemma}
\begin{proof}
We process the clusters $U_1,\ldots,U_n$ in $n$ serial ABP components in order, using DP rules similar to Section~\ref{sec:baseline}.
As unused leaves are interchangeable, our DP only tracks $(a;k,\boldsymbol\nu)$, where $a\in \mathbb N$ counts unused leaves, $k \in \mathbb N$ is the size of the currently open star, and counters $\nu_s$ for $s\in\cR$ and $s\geq r$ count how many stars have grown past size $s$, including the open star.
Let $\boldsymbol\delta_k$ be the unit vector at coordinate $k$ if
$k\in\cR$ and $k\geq r$, and zero otherwise.
\textsc{Begin} takes the smallest unused leaf, while \textsc{Grow}
has $a$ choices, each of weight $-1$; the label $-a$ below abbreviates $a$ parallel transitions.
\begin{alignat}{3}
 &\dplabel{Begin}
 &(a;0,\boldsymbol{\nu})
   &\abparrow{1}(a-1;1,\boldsymbol{\nu}),
 &\qquad a&>0;
 \label{eq:star-formation-begin}\\
 &\dplabel{Grow}
 &(a;k,\boldsymbol{\nu})
   &\abparrow{-a}(a-1;k+1,\boldsymbol{\nu}+\boldsymbol\delta_k),
 &\qquad a,k&>0;
 \label{eq:star-formation-grow}\\
 &\dplabel{Close}
 &(a;k,\boldsymbol{\nu})&\abparrow{1}(a;0,\boldsymbol{\nu}),
 &\qquad k&>r.
 \label{eq:star-formation-close}
\end{alignat}

Use unit transitions to enter $(d_1;0,\mathbf0)$ from $\tau_{\mathbf d}$
and to advance from $(0;0,\boldsymbol\nu)$ in component $j<n$ to
$(d_{j+1};0,\boldsymbol\nu)$ in the next component.
After the last cluster, connect to $\mathsf t_r$ by a unit transition
precisely when $\nu_r\leq\alpha_r n$ and $\nu_s>\alpha_s n$ for every
$s\in\cR$ with $s>r$.
We obtain a topological order along components and decreasing $a$, and taking open states before closed states.

The \textsc{Begin} rule orders the stars within each cluster and fixes
the first leaf of each star. A block $B$ of size $b$ has $(b-1)!$
leaf orders, each of weight $(-1)^{b-1}$, giving its factor $\mu(B)$
as in Lemma~\ref{lem:baseline}.
The final counters satisfy $\nu_s=q_s(\rho)$, so the sink condition
selects exactly the partitions with $r(\rho)=r$.
Summing their weights gives $C_r(\mathbf d)$.
There are $O(n^{|\cR|+3})$ nodes and $O(n)$ outgoing transitions per
node after expanding the $-a$ labels, giving the claimed size bound.
\end{proof}

Now attach a gadget from Lemma~\ref{lem:star-formation-abp} at each closed
terminal $\tau_{\mathbf d}$, with common sink $\mathsf t_r$.
Denote the resulting ABP by $\mathcal B_r$ and its source by
$\mathsf s_r=(\varnothing,\mathbf0;0,0)$.
The attached gadgets multiply each main-part path weight by $C_r(\mathbf d)$,
giving the output
\begin{equation}\label{eq:cutoff-output}
 F_r(A)=D_{\mathcal B_r}[\mathsf t_r]=
 \sum_{\substack{\pi\in\cP_n\\r(\pi)=r}}
       \mu(\pi)\sum_{\varphi\in\Phi(M/\pi)}w(\varphi).
\end{equation}
Indeed, expanding $C_r(\mathbf d)$ pairs the small blocks $\sigma$ from
the main part with large blocks $\rho$; Fact~\ref{fact:large-block-assignment}
ensures that $\pi=\sigma\cup\rho$ is assigned to $r$.
Conversely, each assigned pair $(\pi,\varphi)$ determines a unique
main-part path for each choice of small-star leaf orders with the
smallest leaf first. These paths are retained because their nursery
support is at most $q_r(\pi)\leq\alpha_r n$.
Summing the leaf orders gives $\mu(\sigma)w(\varphi)$ as in
Lemma~\ref{lem:baseline}, including all factors $A_{ij}$; the term
$\rho$ in $C_r(\mathbf d)$ supplies $\mu(\rho)$.
Their product is $\mu(\pi)w(\varphi)$.

\begin{corollary}[Combined ABP]\label{cor:combined-abp}
Suppose every partition of $[n]$ has an admissible cutoff in $\cR$.
Take disjoint copies of $\mathcal B_r$ for $r\in\cR$ and add a new
source $\mathsf s$, a new sink $\mathsf t$, and unit transitions
$\mathsf s\to\mathsf s_r$ and $\mathsf t_r\to\mathsf t$ for each $r$.
The resulting \emph{combined ABP} $\mathcal B$ computes $\per(A)$.
\end{corollary}
\begin{proof}
By \eqref{eq:cutoff-output}, each partition contributes exactly once to
the output $\sum_{r\in\cR}F_r(A)$, which equals $\per(A)$ by
Theorem~\ref{thm:Binet--Minc}.
\end{proof}

\section{The Economics of Stellar Nurseries}\label{sec:parameters}

We now choose the cutoffs and large-star budgets. The budgets must be
small enough that each cutoff ABP retains an exponential saving, yet
generous enough that every partition has an admissible cutoff. The following
parameter lemma shows that both requirements can be met by a fixed finite
choice, independent of $n$.
Throughout, denote binary entropy by
\[
\entropy(\alpha)=-\alpha\log_2\alpha-(1-\alpha)\log_2(1-\alpha).
\]
We will write $\alpha_j$, $c_j$, and $E_j$ for the parameters at cutoff $r_j$.

\begin{lemma}\label{lem:parameters}
There are cutoffs $\cR=\{r_1<\cdots<r_\ell\}$ and rational
parameters $0<\alpha_j<1/2$, $1<c_j<2$, satisfying
$c_j(c_j-1)^{r_j-1}\geq1$ and $E_j<1$ for every $j\in[\ell]$,
such that every partition $\pi\in\cP_n$ has an admissible cutoff for every $n\geq0$.
Here, $E_j=\log_2c_j+\entropy(\alpha_j)-\alpha_j\log_2(c_j-1)$.
\end{lemma}

We first quantify the saving provided by a covered prefix, as in Section~\ref{sec:small-stars} and \cite[Lemma~3]{koivisto}: As the cutoff $r$ grows, larger stars
can cover more leaves without beginning another star, so this restriction
becomes weaker. Nevertheless, in our setup, it still subtracts $\Theta(n/r)$ from the exponent. Eventually, we will show that the  bookkeeping overhead for stellar nurseries fits into this saving. 

\begin{lemma}\label{lem:subset}
Let $N\geq0$ and $r\geq1$ be integers. If $1<c<2$ and
$c(c-1)^{r-1}\geq1$, then
\begin{equation}\label{eq:subset}
 \bigl|\{S\subseteq[N]:|S|\leq r\pref(S)\}\bigr|
                         \leq(N+1)c^N.
\end{equation}
\end{lemma}

\begin{proof}
Fix $p\in\{0,\ldots,N\}$. If $\pref(S)=p$ and $|S|\leq rp$, then
$S=[p]\cup T$ with $T\subseteq\{p+1,\ldots,N\}$ and $|T|\leq(r-1)p$.
Since $0<c-1<1$, each such $T$ has weight
$(c-1)^{|T|-(r-1)p}\geq1$. Their number is therefore at most the sum
of these weights over all subsets $T$, which the binomial theorem evaluates as
\begin{equation}
 \sum_{T\subseteq\{p+1,\ldots,N\}}(c-1)^{|T|-(r-1)p}
 = (c-1)^{-(r-1)p}c^{N-p} = \frac{c^N}{\bigl(c(c-1)^{r-1}\bigr)^p}\leq c^N.
\end{equation}
Summing over the $N+1$ possible values of $p$ proves the claim.
\end{proof}

Recall that the nursery entry $d_j$ counts deferred leaves with chosen
center image $v_j$. Every deferred leaf increases one such count and
extends the covered prefix. The resulting reduction in possible covered
sets can compensate for the number of ways to choose these counts.
We therefore count stellar nurseries together with their compatible
covered sets: even many deferred leaves can be affordable when they
are assigned to few center images.

\begin{proposition}[Effective size at one cutoff]\label{prop:vertices}
For $r\in\cR$, choose $1<c_r<2$ with $c_r(c_r-1)^{r-1}\geq1$.
The cutoff ABP $\mathcal B_r$ has effective size at most
\begin{equation}\label{eq:early}
                 \Ostar*{c_r^n\sum_{p=0}^{\floor{\alpha_r n}}\binom np(c_r-1)^{-p}}.
\end{equation}
\end{proposition}
\begin{proof}
First fix $\mathbf d$ with $m=\sum_jd_j\leq n$.
A path from the source to $(S,\mathbf d;k,j)$ uses exactly $m$
\textsc{Defer} transitions; let $b$ be its number of \textsc{Begin}
transitions. Each of these $m+b$ transitions covers the smallest
uncovered label, and each begun star has at most $r$ leaves. Thus
\begin{equation}\label{eq:nursery-prefix}
 \pref(S)\geq m+b,\qquad |S|-m\leq rb.
\end{equation}
It follows that $[m]\subseteq S$, and the injective map
$S\mapsto T=\{i-m:i\in S\setminus[m]\}\subseteq[n-m]$ gives
$|T|=|S|-m\leq r(\pref(S)-m)=r\pref(T)$.
Then Lemma~\ref{lem:subset}, with $N=n-m$ and $c=c_r$, bounds the number
of such sets $T$, and hence of sets $S$, by
$(n-m+1)c_r^{n-m}\leq(n+1)c_r^{n-m}$.

Next sum over $\mathbf d$. A positive coordinate contributes
$\sum_{a\geq1}c_r^{-a}=1/(c_r-1)$, so choosing its support gives
\begin{equation}\label{eq:endpoint-sum}
 \sum_{\substack{\mathbf d\in\NN^n\\|\supp(\mathbf d)|\leq\alpha_r n}}
             c_r^{-\sum_jd_j}
       =\sum_{p=0}^{\floor{\alpha_r n}}\binom np(c_r-1)^{-p}.
\end{equation}
Extending the sum beyond $m\leq n$ only increases the bound.
Multiplying by $(n+1)c_r^n$ therefore bounds the choices of
$(S,\mathbf d)$. There are at most $1+n\min(r,n)$ choices of $(k,j)$
and $O(n)$ outgoing transitions per node in the main part.
By Lemma~\ref{lem:star-formation-abp}, each attached star-formation gadget has
size $O((n+1)^{|\cR|+4})$. Since $\cR$ is fixed, attaching one per
reachable closed terminal preserves \eqref{eq:early}.
\end{proof}

The binomial entropy bound gives
\begin{equation}\label{eq:entropy}
 \sum_{p=0}^{\floor{\alpha n}}\binom np(c_r-1)^{-p}
 \leq(n+1)2^{n(\entropy(\alpha)-\alpha\log_2(c_r-1))},
 \qquad 0\leq\alpha\leq1/c_r.
\end{equation}
Indeed, $\binom np\leq2^{n\entropy(p/n)}$, and
$\entropy(u)-u\log_2(c_r-1)$ is increasing on $[0,1/c_r]$.
For $r=r_j$ and $\alpha_j\leq1/c_j$, \eqref{eq:early} therefore gives effective size
$\Ostar{2^{E_j n}}$, with $E_j$ as defined in Lemma~\ref{lem:parameters}.
This identifies the main remaining expense, namely, choosing the
occupied center images. For small $\alpha_r$, this costs roughly
$n\alpha_r\log(1/\alpha_r)$ in the exponent. Comparing this with the
saving of order $n/r$ suggests budgets of order $1/(r\log(r+1))$,
with a sufficiently small constant.

Finally, the coverage of partitions relies on the distribution of stars across cutoffs.
A star of size $b$ contributes weight $r_j$ at each cutoff $r_j<b$.
By choosing geometrically increasing cutoffs, we can bounds the sum of these weights by $O(b)$.
For geometrically increasing cutoffs, taking $\alpha_j$ of order
$1/(r_j\log(r_j+1))$ gives weighted thresholds $r_j\alpha_j n$ of
order $n/j$, since $\log(r_j+1)=\Theta(j)$.
Their harmonic sum eventually exceeds the $O(n)$ bound on the total
contribution of all stars, forcing some cutoff test to succeed. The final parameter lemma formalizes this.

\begin{proof}[Proof of Lemma~\ref{lem:parameters}]
Take $r_j=2^{j-1}$, $c_j=2-1/(2r_j)$, and
$\alpha_j=1/(64r_jj)$.
Bernoulli's inequality shows that
$c_j(c_j-1)^{r_j-1}\geq1+(3r_j-1)/(4r_j^2)>1$, and
$\alpha_j<1/2<1/c_j$.
For $r=r_j$, use $\entropy(x)\leq x\log_2(e/x)$,
$-\log_2(c_j-1)\leq1$, $\log_2r=j-1$, and $\log_2j\leq j$ to obtain
\[
\begin{aligned}
 E_j
 &\leq1-\frac1{4r\ln2}
       +\frac{j+6+\log_2j+\log_2e}{64rj} \leq1-\frac1{4r\ln2}+\frac{8+\log_2e}{64r}<1.
\end{aligned}
\]
By harmonic divergence, choose a fixed finite $\ell$ with
$\sum_{j=1}^{\ell}1/j\geq128$.
For every partition $\pi\in\cP_n$,
\begin{equation}\label{eq:dyadic-mass}
 \sum_{j=1}^{\ell}r_jq_{r_j}(\pi)
 =\sum_{B\in\pi}\ \sum_{j:r_j<|B|}r_j
 \leq2\sum_{B\in\pi}|B|=2n,
\end{equation}
since the sum of the dyadic cutoffs below $|B|$ is less than $2|B|$.
If no cutoff were admissible for $\pi$, then $q_{r_j}(\pi)>\alpha_jn$ for every $j$,
and hence we contradict \eqref{eq:dyadic-mass} by $\sum_{j=1}^{\ell}r_jq_{r_j}(\pi)
 >\frac n{64}\sum_{j=1}^{\ell}\frac1j\geq2n.$
\end{proof}

\section{Finishing the Proof}
We have collected all ingredients to prove the main theorem.

\begin{proof}[Proof of Theorem~\ref{thm:main}]
Choose $\cR$ and the parameters from Lemma~\ref{lem:parameters}.
By Corollary~\ref{cor:combined-abp}, the combined ABP $\mathcal B$ computes $\per(A)$.
Let $\gamma=\max_{j\in[\ell]}E_j<1$.
Summing the bounds for $\mathcal B_{r_j}$ from Proposition~\ref{prop:vertices}
and \eqref{eq:entropy} over the fixed finite set $\cR$ gives effective
size $\Ostar{2^{\gamma n}}$ for $\mathcal B$.
Exploring allowed transitions via BFS constructs its reachable part
on input $n \in \mathbb N$ and the fixed parameters, with polynomial work per node. Splitting each $-A_{ij}$ transition into
labels $-1$ and $A_{ij}$ gives elementary labels with constant overhead.
Finally, evaluation via DP in topological order uses linearly many ring operations
in the size of this ABP, so construction and evaluation take
$\Ostar{2^{\gamma n}}$ time and ring operations, respectively.

Setting $\varepsilon=2-2^\gamma>0$ gives a fixed exponential saving.
Using the parameters of Appendix~\ref{app:certificate} in this construction
gives $2^\gamma<2-10^{-40}$ and hence the stated value for $\varepsilon$.
\end{proof}

Finally, by clearing denominators and working over $\ZZ/2^t\ZZ$ with $t \in \mathbb N$ sufficiently large to recover the permanent exactly, we obtain the following bit complexity version of Theorem~\ref{thm:main}.

\begin{corollary}\label{lem:bits}
For $\varepsilon$ as above, there is an algorithm for computing permanents of
$n\times n$ rational matrices with $B$-bit numbers in
$\operatorname{poly}(B) \cdot \Ostar{(2-\varepsilon)^n}$ bit operations.
\end{corollary}

\appendix
\section{Explicit Choice of Constants}\label{app:certificate}

This appendix gives explicit rational parameters for 280 cutoffs
that yield $\varepsilon=10^{-40}$ in the main construction.
For $\cR=\{1=r_1<\cdots<r_\ell\}$ and fractions
$\alpha_j$, set $\beta_1=2$ and $\beta_j=r_j-r_{j-1}$ for
$j\geq2$. Every
partition satisfies
\begin{equation}\label{eq:mass}
 \sum_{j=1}^\ell\beta_jq_{r_j}(\pi)
   =\sum_{B\in\pi}\ \sum_{j:r_j<|B|}\beta_j
   \leq\sum_{B\in\pi}|B|=n.
\end{equation}
For a singleton the inner sum is zero. Otherwise, if $r_i$ is the
largest selected cutoff below $|B|$, it telescopes to $r_i+1\leq|B|$.
Consequently every partition has an admissible cutoff as soon as
\begin{equation}\label{eq:L}
                         L:=\sum_{j=1}^\ell\beta_j\alpha_j\geq1.
\end{equation}
Indeed, if no cutoff were admissible for $\pi$, then
$q_{r_j}(\pi)>\alpha_jn$ for every $j$,
contradicting \eqref{eq:mass}.

\begin{proposition}\label{lem:parameter-certificate}
Take
\begin{equation}\label{eq:parameters}
 \ell=280,\qquad r_1=1,\qquad
 r_{j+1}=\left\lceil\frac{4r_j}{3}\right\rceil\quad(1\leq j<\ell),
\end{equation}
and set $K=r_\ell<10^{36}$. Use
\begin{equation}\label{eq:explicit-thresholds}
 \begin{aligned}
 &\alpha_1=\frac{11}{100},\qquad c_1=\frac{100}{89},\\
 &\alpha_j=\frac1{r_j(j+13)},\qquad
 c_j=2-\frac2{3r_j}\quad(2\leq j\leq\ell).
 \end{aligned}
\end{equation}
These parameters satisfy
\[
 L>1.001,\qquad E_j<1-10^{-38}\quad(1\leq j\leq\ell),
\]
as well as $c_j(c_j-1)^{r_j-1}>1$ and
$0<\alpha_j<1/c_j$. In particular, one may take $\varepsilon=10^{-40}$ in the $\Ostar{(2-\varepsilon)^n}$ bounds for ABP size, construction time, and ring operations.
\end{proposition}

\subsection{Bounds for the Parameters}

For rational $1\leq y\leq2$, put $z=(y-1)/(y+1)$ and
\[
 P(y)=2\sum_{h=0}^{63}\frac{z^{2h+1}}{2h+1},\qquad
 R(y)=\frac{2z^{129}}{129(1-z^2)}.
\]
The expansion $\ln y=2\sum_{h\geq0}z^{2h+1}/(2h+1)$ gives
\[
                    P(y)\leq\ln y\leq P(y)+R(y).
\]
For any positive rational $x$, write $x=2^ky$ with $1\leq y<2$.
Adding the intervals for $\ln y$ and $k\ln2$ (reversing the latter
endpoints when $k<0$) gives rational bounds
$\lambda_-(x)\leq\ln x\leq\lambda_+(x)$.

For every triple $(r_j,\alpha_j,c_j)$ in
\eqref{eq:parameters}--\eqref{eq:explicit-thresholds}, the following
rational inequalities hold:
\begin{align*}
 &1<c_j<2,\qquad 0<\alpha_j<\tfrac12<1/c_j,\\
 &\lambda_-(c_j)+(r_j-1)\lambda_-(c_j-1)>0,\\
 &0<U_j<1-10^{-38},
\end{align*}
where
\[
 U_j=
 \frac{\lambda_+(c_j)
 -\alpha_j\lambda_-(\alpha_j)
 -(1-\alpha_j)\lambda_-(1-\alpha_j)
 -\alpha_j\lambda_-(c_j-1)}{\lambda_-(2)}.
\]
The logarithmic inequality proves the subset-counting condition. The
positive numerator and $\lambda_-(2)>0$ give $E_j\leq U_j$, proving
the required effective-size bounds for the cutoff ABPs by Section~\ref{sec:parameters}.
The weighted mass is the rational number
\[
 L=\frac{11}{50}
   +\sum_{j=2}^{280}\frac{r_j-r_{j-1}}{r_j(j+13)}.
\]
Direct rational summation gives
\[
 1.001332419877\leq L<1.001332419878.
\]
Thus \eqref{eq:mass} guarantees that every partition meets at least
one cutoff test, for every $n$.
Finally, $r_{j+1}+3\leq\frac43(r_j+3)$ implies
$K=r_{280}\leq4(4/3)^{279}-3<10^{36}$.
Convexity gives $2^{1-10^{-38}}\leq2-10^{-38}<2-10^{-40}$, proving the claimed choice of $\varepsilon$.

\clearpage
\section{Definitions for the Lean Statement}\label{app:lean}

We expand the definitions used by \texttt{main\_abp} on page~\pageref{lst:lean-main}.
\begin{figure}[H]
\centering
\lstset{style=lean,xleftmargin=0pt,xrightmargin=0pt,
        aboveskip=0pt,belowskip=0pt}
\newsavebox{\leanDepMain}
\newsavebox{\leanDepLabel}
\newsavebox{\leanDepABP}
\newsavebox{\leanDepEval}
\newsavebox{\leanDepPaths}
\newsavebox{\leanDepOutput}
\newsavebox{\leanDepPermanent}
\begin{lrbox}{\leanDepMain}\begin{minipage}{.97\linewidth}
\input{figures/lean-abp-statement.tex}
\end{minipage}\end{lrbox}
\begin{lrbox}{\leanDepLabel}\begin{minipage}{.44\linewidth}
\begin{lstlisting}
inductive Label (n : ℕ)
  | variable (i j : Fin n)
  | constant (c : ℤ)
\end{lstlisting}
\end{minipage}\end{lrbox}
\begin{lrbox}{\leanDepABP}\begin{minipage}{.44\linewidth}
\begin{lstlisting}
structure ABP (n : ℕ) where
  (vertices transitions : ℕ)
  (source sink : Fin vertices)
  (tail head :
    Fin transitions → Fin vertices)
  label : Fin transitions → Label n
  rank : Fin vertices → Fin vertices
  increasing : ∀ e,
    rank (tail e) < rank (head e)
\end{lstlisting}
\end{minipage}\end{lrbox}
\begin{lrbox}{\leanDepEval}\begin{minipage}{.44\linewidth}
\begin{lstlisting}
def Label.eval : Label n → ℤ
  | .variable i j => A i j
  | .constant c => c
\end{lstlisting}
\end{minipage}\end{lrbox}
\begin{lrbox}{\leanDepPaths}\begin{minipage}{.54\linewidth}
\begin{lstlisting}
noncomputable def ABP.paths (G : ABP n) :
    ℕ → Fin G.vertices → ℤ
  | 0, v => if v = G.source then 1 else 0
  | k+1, v =>
    (if v = G.source then 1 else 0) +
    ∑ e : Fin G.transitions,
      if G.head e = v then
        (G.label e).eval A * G.paths k (G.tail e)
      else 0
\end{lstlisting}
\end{minipage}\end{lrbox}
\begin{lrbox}{\leanDepOutput}\begin{minipage}{.44\linewidth}
\begin{lstlisting}
noncomputable def ABP.output (G : ABP n) :=
  G.paths A G.vertices G.sink
\end{lstlisting}
\end{minipage}\end{lrbox}
\begin{lrbox}{\leanDepPermanent}\begin{minipage}{.44\linewidth}
\begin{lstlisting}
noncomputable def permanent : ℤ :=
  ∑ p : Equiv.Perm (Fin n),
    ∏ i, A i (p i)
\end{lstlisting}
\end{minipage}\end{lrbox}
\begin{lstlisting}[aboveskip=0pt,belowskip=8pt]
variable {n : ℕ} (A : Fin n → Fin n → ℤ)
\end{lstlisting}
\begin{tikzpicture}[diagram,
 lean dependency/.style={draw=black!18,fill=black!1,
   rounded corners=2pt,inner sep=5pt,outer sep=1pt},
 lean arrow/.style={diagram edge,->,shorten <=2pt,shorten >=2pt}]
 \node[lean dependency] (main) {\usebox{\leanDepMain}};
 \node[lean dependency,anchor=north] (output)
   at ([xshift=-.25\linewidth,yshift=-8mm]main.south)
   {\usebox{\leanDepOutput}};
 \node[lean dependency,anchor=north] (permanent)
   at ([xshift=.25\linewidth,yshift=-8mm]main.south)
   {\usebox{\leanDepPermanent}};
 \node[lean dependency,anchor=north] (paths)
   at ([yshift=-8mm]main.center |- permanent.south)
   {\usebox{\leanDepPaths}};
 \node[lean dependency,anchor=north] (abp)
   at ([xshift=-.25\linewidth,yshift=-8mm]paths.south)
   {\usebox{\leanDepABP}};
 \node[lean dependency,anchor=north] (eval)
   at ([xshift=.25\linewidth,yshift=-8mm]paths.south)
   {\usebox{\leanDepEval}};
 \node[lean dependency,anchor=north] (label)
   at ([yshift=-8mm]main.center |- abp.south)
   {\usebox{\leanDepLabel}};

 \draw[lean arrow] ([xshift=-.25\linewidth]main.south) -- (output.north);
 \draw[lean arrow] ([xshift=.25\linewidth]main.south) -- (permanent.north);
 \draw[lean arrow] (output.south) -- (paths.north);
 \draw[lean arrow] (paths.south) -- (abp.north);
 \draw[lean arrow] (paths.south) -- (eval.north);
 \draw[lean arrow] (abp.south) --
   ([xshift=-13mm]label.north);
 \draw[lean arrow] (eval.south) --
   ([xshift=13mm]label.north);
\end{tikzpicture}
\label{fig:lean-abp-dependencies}
\end{figure}

\end{document}